\documentclass[amsmath,amssymb,
 preprint%
]{revtex4-1}

\usepackage{color}
\usepackage{graphicx}% Include figure files
\usepackage{dcolumn}% Align table columns on decimal point
\usepackage{bm}% bold math
\begin{document}

\preprint{AIP/123-QED}

\title{Optical phase noise engineering via acousto-optic interaction and its interferometric applications\\}

%% For REVTeX it is possible to automate superscript and e-mail callouts with the superscriptaddress option; see REVTeX4 documentation.

\author{Nandan Satapathy}
 \email{nandan.s@gmail.com}
  \affiliation{ Raman Research Institute, C.V. Raman Avenue, Sadashiva Nagar, Bangalore, INDIA-560080%  \\This line break forced with \textbackslash\textbackslash
}%

\author{Deepak Pandey}  
 \affiliation{ Raman Research Institute, C.V. Raman Avenue, Sadashiva Nagar, Bangalore, INDIA-560080%  \\This line break forced with \textbackslash\textbackslash
}% 

\author{Sourish Banerjee}
 \altaffiliation[Currently at ]{Department ME/EI, TU Delft, The Netherlands }%Lines break automatically or can be forced with \\
  \affiliation{ Raman Research Institute, C.V. Raman Avenue, Sadashiva Nagar, Bangalore, INDIA-560080%  \\This line break forced with \textbackslash\textbackslash
}%
     
\author{Hema Ramachandran}
 \email{hema@rri.res.in}
  \affiliation{ Raman Research Institute, C.V. Raman Avenue, Sadashiva Nagar, Bangalore, INDIA-560080%  \\This line break forced with \textbackslash\textbackslash
}%

\begin{abstract}
We exercise  rapid and fine control over the phase of light by transferring 
digitally generated phase jumps from radio frequency (rf) electrical signals onto light by means 
of  acousto-optic interaction. By tailoring the statistics of phase jumps in 
the electrical signal and thereby  engineering the optical phase noise, we 
manipulate  the visibility of  interference fringes  in  a Mach-Zehnder 
interferometer that incorporates two acousto-optic modulators. Such controlled 
dephasing finds applications in modern experiments involving the spread or 
diffusion of light in an optical network. Further, we analytically show how 
engineered partial phase noise can convert the dark port of a stabilised 
interferometer to a weak source of highly correlated photons. 
\end{abstract}


\maketitle

\noindent 

The fine control of optical phase shifts is of prime importance in numerous 
physical applications. On the one hand, interferometric measurements in fields 
as diverse as holography and gravitational wave detection demand precise 
compensation of random phase fluctuations arising out of environmental 
disturbances. On the other hand, quantum information and state manipulation 
demand deterministic phase shifts that are  rapid and precise. A
requirement that has recently emerged is that of controlled {\it dephasing}, 
critical in applications like optical implementation of quantum walks\cite{Broome, Schreiber, 
Pandey} and optical simulation of noise-assisted coherent transport\cite{Caruso} 
that seek to enhance the spread or diffusion of light in a network. 

Light may be dephased by introducing random phase shifts in a variety of ways.
An increase in the optical path length  resulting in the phase shift of the emergent light may be effected on the milliseconds time scale, by means of the piezo-mechanical movement of an 
optical element or the use of liquid-crystal spatial light modulators and retarders. 
Recent quantum state manipulation experiments utilise  electro-optic 
modulators \cite{Klinger} that typically respond in the $\sim 10\mu$s  timescale or less;  
these not only are polarisation-sensitive  but also  have a limited range. An 
acousto-optic modulator (AOM) too may  be used as a phase shifter, as shown by Li 
{\it et al} \cite{Li} with a  resolution  of $6^\circ$,  by Sadgrove {\it et al} 
\cite{Sadgrove} who imparted phase changes on time scales $\approx 30 \mu$s and by 
Pandey {\it et al}, who utilised its  polarisation insensitivity \cite{Pandey}. 

In this Letter we use appropriately tailored radio-frequency (rf) electrical  input to impart 
random phase jumps to light via the acousto-optic interaction, at time intervals as 
short as 500ns and with a phase resolution of $0.01^\circ$, thereby engineering  any desired 
optical phase noise. We begin this Letter with a  simple analysis of acousto-optically 
induced optical phase transfer from radio-frequency electrical signal, followed by its experimental 
demonstation in a Mach-Zehnder interferometer, that reveals the  transfer of 
different phase shifts  to the different orders of diffraction. The  control of the visibility of fringes is demonstrated by  using light with 
different phase noise statistics. Some possible applications are discussed. Finally, 
we analytically show  how engineered partial phase noise can convert the dark port of 
a stabilised  interferometer to a weak source of highly correlated, or bunched, photons.

In an AOM, an  externally applied rf electrical signal of frequency $\Omega_{rf}$ 
creates an acoustic strain wave of the same frequency and of wavelength 
$\Lambda_{s}$ \cite{Saleh-Teich}. The accompanying  refractive index change, 
proportional to the amplitude of the strain, thus creates a moving grating 
that diffracts light of wavelength $\lambda$ and frequency $\omega$ according to 
the grating equation, which in the Fraunhoffer limit is 
$\Lambda_{s}(sin\theta_{i}+ sin\theta_{n})=n\lambda$, where $\theta_{i}$ 
and $\theta_{n}$ are the angles of incidence and diffraction of light, and $n$ is the 
order of diffraction. As the grating is moving with velocity 
$V_{s}=\Omega_{rf}\Lambda_{s}$, the $n^{th}$ order diffracted light suffers a doppler 
shift and has a resulting frequency $\omega_{n}=\omega(1+n(V_{s}/\Lambda_{s})(\lambda/c))
=\omega+n\Omega_{rf}$. Thus the frequency of the diffracted light in the different orders 
are shifted by integral multiples of the applied radio frequency. The angle of 
diffraction  is determined by the wavelength of light and that of the acoustic wave, 
while the diffraction efficiency is determined by the rf power. A phase shift of 
$\phi_{rf}$ to the acoustic wave would effectively shift the grating in space by an 
amount $\delta_s = \Lambda_s\phi_{rf}/2\pi$ leading to a change in the path length of 
the $n^{th}$ order diffracted beam by
\begin{eqnarray} 
\delta_{\lambda n} = (sin\theta_{i}+sin\theta_{n})\delta_{s} = n \lambda \phi_{rf}/2\pi
\end{eqnarray}
The corresponding  phase shift  $\phi_n$ imparted to the $n^{th}$ order diffracted light beam is 
$n\phi_{rf}$. It may be noted that  different orders acquire different phase shifts, 
that may be positive or negative, depending on the geometry. 

\begin{figure}
\center
\includegraphics[width = 8.5cm]{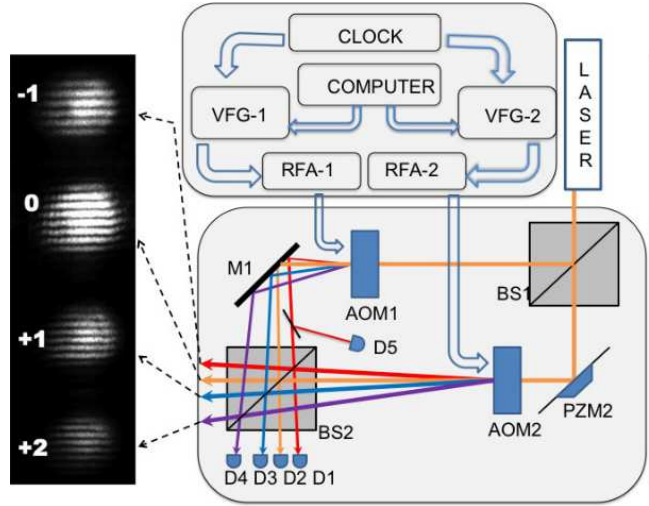}
%[scale=0.25,angle=0]%[width=8.3cm]{AOM_setup_OL.jpg}%height=5.5cm, 
\caption{(Color online) Schematic of the experimental setup}
\label{AOM_setup}
\end{figure}

As changes in phase of light can be seen only in interference, a Mach-Zehnder 
interferometer, with an AOM inserted in each of the two paths of the interferometer, 
was used to study the transfer of phase jumps from the  rf signal to light via the acousto-optic 
interaction. Light from an external cavity diode laser at 767nm  was divided into two 
equal parts at BS1 (Fig.\,\ref{AOM_setup}). These were incident on acousto-optic modulators 
(Isomet) AOM1 and AOM2 that were aligned such that the light intensity was distributed into the 
diffraction orders $ n = -1,0,1,2$. The corresponding orders of the diffracted light from 
the two arms of the interferometer were combined at  beamsplitter BS2. Light emerging 
from one exit port was incident on a screen where the interference patterns of all the 
four orders could be viewed. Light emerging from the other port was made incident on 
photodetectors $D1$ to $D4$ that measured the resultant intensities of the interfering 
diffracted beams of different orders, at a given fringe position. Digital frequency 
synthesizers (Toptica) VFG1 and VFG2 operating at 80 MHz, followed by radio-frequency 
amplifiers RFA1 and RFA2, were used to drive the two AOMs. In order to obtain stable 
fringes the digital rf signal generators were frequency locked using a 10MHz reference 
clock signal. The digital rf signal generator, based on DDS (direct digital synthesizer) 
technology in a FPGA (frequency programmable gate array) platform, has a frequency 
resolution of $\sim 50mHz$ and a phase resolution of $\sim 0.1mrad$, and permits 
changes on a time scale of $500ns$ (sustained) and even $5ns$ (brief). The entire 
experiment was set up on a vibration isolation optical table and shielded from 
air-drafts. Interference patterns were stable over $\sim1s$. Mirror $M1$ and piezo 
operated mirror $PZM2$ were adjusted such that line fringes were seen on all the orders. 
Detectors $D1$ to  $D4$ had active areas smaller than a fringe width, so that  any shift of 
a fringe, say by the introduction of an additional phase shift, was recorded as a change 
in the detector signal.
\begin{figure}
\center
\includegraphics
[width=8.5cm]{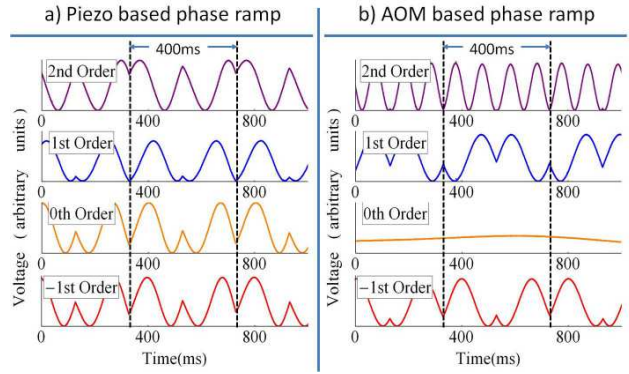}
\caption{(Color online) Detector recordings of the intensity of interference at a fixed location for 
various orders of diffracted light when (a) PZM2 is oscillated at $2.5Hz$, and (b) the 
phase of the rf is ramped from $0^\circ$ to $360^\circ$ at 2.5Hz.}
\label{AOM_pzt}
\end{figure}

In contrast to mechanical disturbances, that cause the same amount of phase shift to the undiffracted 
light and light that is diffracted to various orders, the acousto-optically imparted 
phase shifts affect the various orders to different extents. Oscillation of mirror $PZM2$ 
by means of a 2.5Hz ramp voltage showed that fringes of all orders oscillate in unison, at 
the same frequency as the ramp voltage and in phase (Fig.\,\ref{AOM_pzt}a). Next %PZM2 is no 
%longer oscillated and a 2.5Hz digital phase ramp from $0^\circ$ to $360^\circ$ is generated 
%(with $10^4$ samples in this range). Using LabVIEW, and communication through a USB2.0, 
VFG1 was programmed to generate an rf signal of $80$MHz with a superposed digital phase ramp from $0^\circ$ to $360^\circ$ ($10^4$ samples) which was then amplified and fed to AOM1,  
while AOM2 received just the $80$MHz rf. The signals recorded on the  photodetectors $D1$ 
to $D4$ (Fig.\ref{AOM_pzt}b) showed that the undiffracted $0^{th}$ order remained 
unaffected; the $1^{st}$ order  oscillated at the same frequency as the phase modulation 
of the rf signal; the $-1^{st}$ order oscillated at the same frequency, but exactly out 
of phase with the $1^{st}$ order beam; and the $2^{nd}$ order oscillated at twice the 
frequency, exactly as expected from the earlier analysis. 

\begin{figure}
\center
\includegraphics
[width=8.5cm]{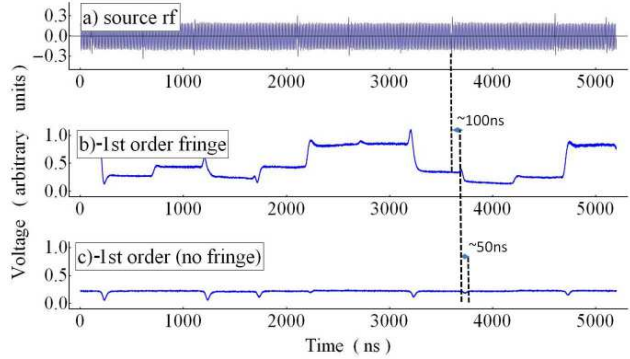}
\caption{(Color online) (a) The rf signal with abrupt phase jumps, that was fed to AOM1, (b) the 
interference signal recorded using the $-1^{st}$ order diffracted light (D1), and (c) the intensity 
of a weak pick-off beam  from the $-1^{st}$ order diffracted light (D5).}
\label{abrupt-jumps}
\end{figure}

This technique of changing the phase of light through acousto-optic interaction makes possible the creation of a light source with phase 
noise alone - a kind of source considered theoretically by Baym\cite{Baym}, and recently 
realized by Satapathy {\it et al} in a different context\cite{Satapathy}. We 
examined this by supplying rf electrical signal with  abrupt phase jumps of random values to 
AOM1 at intervals of 500ns (Fig. \ref{abrupt-jumps}(a)). 
These resulted in  abrupt fringe shifts causing a sudden change in the detector signal 
level from one quasi-stable value to another (Fig.\,\ref{abrupt-jumps}(b)). The single 
beam intensity measured by detector D5 prior to the interference, however, remained 
essentially constant (Fig. \ref{abrupt-jumps} (c)), 
confirming that a phase change to the rf alters only the phase of the diffracted light and 
not its intensity.  The small glitches ($\sim 50ns$; limited by bandwidth of our 
photodetector) seen both in the single beam intensity and the two-beam interference 
signals arise predominantly due to the fact that at the instant of the abrupt phase 
changes, the frequency spread of  the rf and consequently  the angular spread of the 
diffracted light is large. The $\sim100ns$ lag observed between the phase jump in the rf 
and the shift in detector signal level is attributed to the slow speed of the acoustic 
strain wave within the AOM crystal. 
  
\begin{figure}
\includegraphics[width=8.5cm]{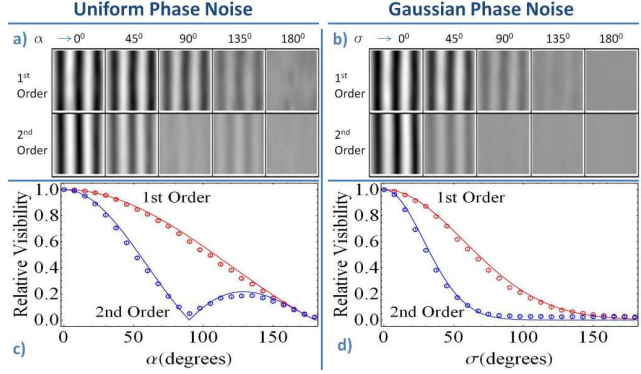}
\caption{(Color online) Interference patterns recorded for the $1^{st}$ and 
$2^{nd}$ order diffracted light for various parameters of 
 (a) uniform, and (b) gaussian 
distributions of phase jumps;  (c) and (d) are the visibilities extracted from the 
data (dots) shown along with the theoretically expected 
curves. The error bars (standard deviation) are mostly within the circles.}
\label{noise}
\end{figure}

By tailoring the amplitude and the statistical distribution of the phase jumps, light with 
different phase noise statistics and varying extents of dephasing  may be created. We use
this technique to demonstrate the control of the visibility of interference fringes. Long 
sequences of computer generated psuedo-random numbers of various statistical distibutions 
of phase jumps (with $0.01^\circ$ resolution) were fed to  VFG1 that was then programmed 
to generate a $80$MHz rf signal with its phase altered in that fashion. The parameters 
amenable to manipulation were the functional form of the probability distribution, 
the mean residence time, and the amplitude of phase excursions. The rf signal was fed to AOM1 
and the interference pattern of the  $1^{st}$ and $2^{nd}$ order diffracted light 
recorded with a CCD camera (Watec, 25 fps). Fig.\,\ref{noise}(a) shows the images recorded 
for a uniform distribution of amplitude of phase jumps, for various standard deviations and 
Fig. 
\ref{noise}(b) the same for a  gaussian distribution. For a perfectly coherent source, 
and equal intensities of the  interfering beams, $V =1$. In the presence of phase noise the 
visibility of $n^{th}$ order fringe $V_n=\left\langle\cos(n\phi)\right\rangle$, where 
$\phi$ is the random rf phase difference between the two AOMs. For a uniform distribution 
of rf phase jumps in range $-\alpha\leq\phi\leq \alpha$,
\begin{eqnarray}
V_{n}(\alpha) = \left|\frac{1}{2\alpha}\int\limits_{-\alpha}^{\alpha}
{\cos(n\phi)\,d\phi}\right|=\left|\frac{\sin(n\alpha)}{n\alpha}\right|
\label{eq-uniform}
\end{eqnarray}
and for a  gaussian distribution of standard deviation $\sigma$,
\begin{eqnarray}
V_{n}(\sigma) = \left|\frac{1}{\sqrt{2\pi}\sigma}\int\limits_{-\infty}^{\infty}
{\cos(n\phi)e^{\frac{-\phi^2}{2\sigma^2}}\,d\phi}\right|=e^{\frac{-n^2\sigma^2}{2}}
\label{eq-gaussian}
\end{eqnarray}
The average relative visibility ($V_n(\alpha \,\text{or}\, \sigma)/V_n(0)$), determined experimentally
from a set of $50$ images of interference patterns of the $1^{st}$ and $2^{nd}$ order diffracted 
light, are shown as circles in Fig.\,\ref{noise}(c) and (d)  for 
various parameters of uniform and 
gaussian phase noises, respectively. They match quite well with 
the theoretical curves obtained from Eqs. \ref{eq-uniform} and 
\ref{eq-gaussian}. This effectively demonstrates the controlled 
dephasing of light using AOMs. In contrast to EOM-based dephasing that suffers from a voltage limit and therefore to a 
phase-jump limit, the use of an AOM does not impose a 
restriction on the amplitude of the phase jump. 

Dephasing or phase noise has traditionally been sought to be 
eliminated.
However, contrary to belief, noise has recently been shown to be 
benificial in certain situations like white noise induced 
entanglement of light \cite{Plenio1, Beige}, 
noise-assisted enhancement of channel capacity in the transmission of classical and quantum information 
\cite{Caruso2} and the more uniform spread of the quantum walker under partial decoherence 
\cite{Kendon}. Currently the role of dephasing in photosynthesis and in increasing the efficiency of solar energy harvesting complexes is under intense 
investigation\cite{Mohseni, Plenio, Caruso3}. 
Underlying the photonic implementations and simulations of these 
phenomena is the 
multistage interference of light \cite{Broome, Schreiber, Pandey, Caruso}. Our work opens up the possibility of 
introducing phase noise of the desired statistics in a controlled fashion which could lead to  the optimization of the 
noise-induced effects in such systems. 

\begin{figure}
\includegraphics[width=8.5cm]{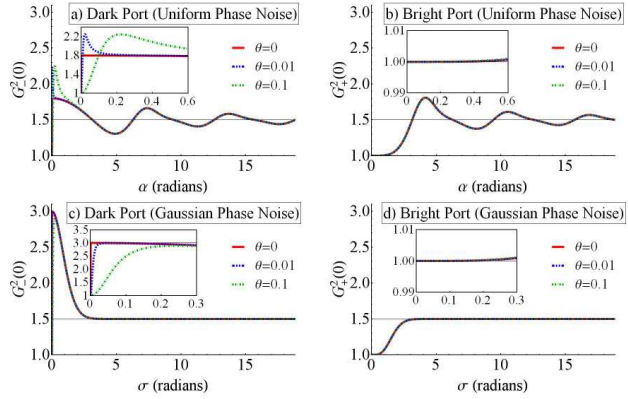} 
\caption{(Color online) Theoretically obtained $G^2(0)$  for various values of 
$\theta$ as function of noise parameter $\alpha$ of uniform 
phase noise for the  (a) dark port and the  (b) bright port of a stabilized 
Mach-Zehnder interferometer; the same for noise parameter $\sigma$ of gaussian phase 
noise for the (c) dark port and the (d) bright port. 
Insets show the expanded view of the plots for $\alpha$ and $\sigma$ close 
to zero.}
\label{g2}
\end{figure}

Using the technique of introducing controlled phase noise in an interferometer, we  arrive 
at an interesting result in a 
completely different context - the creation of a weak source of highly correlated photons. 
Let us consider again the interferometer of Fig. 1 that has been perfectly aligned and 
stabilized with a phase difference $\theta\sim 0 $ between two arms, such that 
one port is nearly dark (-) and the other port bright (+). With phase noise 
introduced in one of the AOMs, instantaneous intensities at the two exit ports would be 
given by $I_{\pm}(t) = I_0[1\pm\cos(\theta+\phi(t)]$ where $I_0$ is the intensity of the first order diffracted light at each AOM. 
The second order correlation or intensity-intensity correlation of the light emerging from 
two exit ports, is $G^2_{\pm}(\tau)= \left\langle I_{\pm}(t)I_{\pm}(t+\tau) 
\right\rangle_t /\left\langle I_{\pm}(t) \right\rangle_t^2$. The zero delay second order 
correlation is,
\begin{eqnarray}
G^2_{\pm}
(0)=1+\frac{\left\langle\cos^2(\theta+\phi(t))\right\rangle_t-\left\langle\cos
(\theta+\phi(t))\right\rangle_t^2}{\left\langle1\pm\cos(\theta+\phi(t))\right\rangle_t^2}
\end{eqnarray}
For uniform phase noise in the range $-\alpha\leq\phi\leq\alpha$ being fed to AOM1, 
\begin{eqnarray}
G^2_{\pm}(0)=1+\frac{\frac{1}{2}\left( \cos (2\theta)\frac{ \sin (2\alpha) }{2 \alpha}+1\right)-\cos^2(\theta)\frac{\sin^2(\alpha)}{\alpha^2}}{1\pm2\cos(\theta)
\frac{\sin(\alpha)}{\alpha}+\cos^2(\theta)\frac{\sin^2(\alpha)}{\alpha^2}}
\end{eqnarray}
while for   gaussian phase noise of standard deviation $\sigma$,
\begin{eqnarray}
G^2_{\pm}(0)=1+\frac{\frac{1}{2}\left(\cos(2\theta)e^{\frac{-2^2\sigma^2}{2}}+1\right)
-\cos^2(\theta)e^{\frac{-2\sigma^2}{2}}}{1\pm2\cos(\theta)e^{\frac{-\sigma^2}{2}}
+\cos^2(\theta)e^{\frac{-2\sigma^2}{2}}}
\end{eqnarray}
For complete phase noise i.e., when $\alpha=n\pi$ or $\sigma=\infty$, we 
see from eqns 5 and 6 that $G^2_{\pm}(0)=1.5$. Similarly in the absence of phase noise  i.e., when $\alpha$ or 
$\sigma$ is $0$, $G^2_{\pm}(0)=1$ for all $\theta\neq 0$. The case $\theta=0$ i.e one port 
completely dark and the other maximally bright, is interesting and needs further elaboration. 
For $\theta=\alpha=\sigma=0$, $G^2_-(0)$ is, mathematically speaking, undefined. However 
this may be evaluated in two ways in the vicinity of $\theta=0$. Fixing $\alpha$ ($\sigma$) 
equals to zero and then taking the limit on $\theta$ one gets the value $G^2_-(0)=1$, the 
value that one obtains for a coherent source. Instead, fixing 
$\theta=0$ and taking the limit 
of $\alpha$ ($\sigma$) tending to zero, one obtains  $G^2_-(0)$ 
equals to $1.8$ ( $3$). The latter  
may be physically understood as the dark port of a well 
stabilized Mach-Zehnder interferometer becoming a  weak 
source of highly correlated or bunched photons 
in the presence of a small engineered phase noise, with the value 
of correlation depending on the nature of the phase noise. 
Interestingly, 
as $G_+^2(0)$ remains nearly $1$ in the presence of small phase noise, there is  negligible 
loss of coherence in the light emerging from the bright port. In  Fig.\,\ref{g2} $G^2_{\pm}(0)$ are plotted 
for various values of $\theta$ close to zero to show how the 
limit is approached. It can be seen that for uniform phase noise $G^2_-(0)$ attains a maximum $(> 2)$ for $\theta$ close to, but not equal to, $0$.  

Recently, higher order intensity correlations have been resorted to for the 
enhancement of contrast in ghost imaging using thermal light\cite{Ou, Bai, Agafanov, Cao}. However, this has a limiting value of 
$N!$ for the $N^{th}$ order correlation; this limit is $2$ for second order intensity correlation\cite{Cao}. As shown in Fig.\,\ref{g2}, $G^2_-(0)$ exceeds 
the value 2 for a range of parameters of uniform and gaussian phase noise. In fact, dramatic 
enhancement results even for the second order correlation if one were to impart phase noise 
having the  Lorentzian (Cauchy) probability 
distribution  function
$\left[ \pi  \gamma  \left(\frac{(x-\theta )^2}{\gamma ^2}+1\right) \right]^{-1}$ , with a noise parameter $\gamma$. For $\theta=0$, we have 
$G_{\pm}(0) =1+\frac{1}{2}\frac{(1\mp e^{-\gamma})}{(1\pm e^{-\gamma})} $. In the limiting value 
of $\gamma\rightarrow0$, $G_-(0)$ diverges, suggesting that very high values of correlation can be obtained with a small amount of engineered Lorentzian phase noise. 

As described earlier, in the interferometer of Fig.\,\ref{AOM_setup}, all orders of diffracted light 
are seen to  suffer the same phase shift on mechanical vibrations, but different amounts due to 
acousto-optic interaction. This could be made use of in introducing rapid and precise partial 
dephasing of light.  For this, the $0^{th}$ order fringe should be used for stabilisation of the 
interferometer against uncontrolled environmental fluctuations by conventional techniques (like feedback to 
the piezo), while engineered phase noise is introduced to the AOM to dynamically dephase the 
diffracted light in a controlled fashion.

To conclude, we have shown 
how an AOM may be used to impart phase noise of desired characteristics to light in a rapid, controlled 
and precise manner. This allows for controlled dephasing of light; an important feature is 
optimizing several light transport based phenomena. We have also shown the possibility of creating 
(weak) sources of highly correlated photons using a balanced Mach-Zehnder interferometer with engineered partial phase noise. The ability to 
control both the interval between phase jumps and the amplitude of phase jumps opens up the possibility of 
creating classical, incoherent light sources with tunable temporal coherence\cite{Pandey2}.


\begin{thebibliography}{22}%
\makeatletter
\providecommand \@ifxundefined [1]{%
 \@ifx{#1\undefined}
}%
\providecommand \@ifnum [1]{%
 \ifnum #1\expandafter \@firstoftwo
 \else \expandafter \@secondoftwo
 \fi
}%
\providecommand \@ifx [1]{%
 \ifx #1\expandafter \@firstoftwo
 \else \expandafter \@secondoftwo
 \fi
}%
\providecommand \natexlab [1]{#1}%
\providecommand \enquote  [1]{``#1''}%
\providecommand \bibnamefont  [1]{#1}%
\providecommand \bibfnamefont [1]{#1}%
\providecommand \citenamefont [1]{#1}%
\providecommand \href@noop [0]{\@secondoftwo}%
\providecommand \href [0]{\begingroup \@sanitize@url \@href}%
\providecommand \@href[1]{\@@startlink{#1}\@@href}%
\providecommand \@@href[1]{\endgroup#1\@@endlink}%
\providecommand \@sanitize@url [0]{\catcode `\\12\catcode `\$12\catcode
  `\&12\catcode `\#12\catcode `\^12\catcode `\_12\catcode `\%12\relax}%
\providecommand \@@startlink[1]{}%
\providecommand \@@endlink[0]{}%
\providecommand \url  [0]{\begingroup\@sanitize@url \@url }%
\providecommand \@url [1]{\endgroup\@href {#1}{\urlprefix }}%
\providecommand \urlprefix  [0]{URL }%
\providecommand \Eprint [0]{\href }%
\providecommand \doibase [0]{http://dx.doi.org/}%
\providecommand \selectlanguage [0]{\@gobble}%
\providecommand \bibinfo  [0]{\@secondoftwo}%
\providecommand \bibfield  [0]{\@secondoftwo}%
\providecommand \translation [1]{[#1]}%
\providecommand \BibitemOpen [0]{}%
\providecommand \bibitemStop [0]{}%
\providecommand \bibitemNoStop [0]{.\EOS\space}%
\providecommand \EOS [0]{\spacefactor3000\relax}%
\providecommand \BibitemShut  [1]{\csname bibitem#1\endcsname}%
\let\auto@bib@innerbib\@empty
%</preamble>
\bibitem [{\citenamefont {Broome}\ \emph {et~al.}(2010)\citenamefont {Broome},
  \citenamefont {Fedrizzi}, \citenamefont {Lanyon}, \citenamefont {Kassal},
  \citenamefont {Aspuru-Guzik},\ and\ \citenamefont {White}}]{Broome}%
  \BibitemOpen
  \bibfield  {author} {\bibinfo {author} {\bibfnamefont {M.~A.}\ \bibnamefont
  {Broome}}, \bibinfo {author} {\bibfnamefont {A.}~\bibnamefont {Fedrizzi}},
  \bibinfo {author} {\bibfnamefont {B.~P.}\ \bibnamefont {Lanyon}}, \bibinfo
  {author} {\bibfnamefont {I.}~\bibnamefont {Kassal}}, \bibinfo {author}
  {\bibfnamefont {A.}~\bibnamefont {Aspuru-Guzik}}, \ and\ \bibinfo {author}
  {\bibfnamefont {A.~G.}\ \bibnamefont {White}},\ }\href {\doibase
  10.1103/PhysRevLett.104.153602} {\bibfield  {journal} {\bibinfo  {journal}
  {Phys. Rev. Lett.}\ }\textbf {\bibinfo {volume} {104}},\ \bibinfo {pages}
  {153602} (\bibinfo {year} {2010})}\BibitemShut {NoStop}%
\bibitem [{\citenamefont {Schreiber}\ \emph {et~al.}(2011)\citenamefont
  {Schreiber}, \citenamefont {Cassemiro}, \citenamefont
  {Poto\ifmmode~\check{c}\else \v{c}\fi{}ek}, \citenamefont {G\'abris},
  \citenamefont {Jex},\ and\ \citenamefont {Silberhorn}}]{Schreiber}%
  \BibitemOpen
  \bibfield  {author} {\bibinfo {author} {\bibfnamefont {A.}~\bibnamefont
  {Schreiber}}, \bibinfo {author} {\bibfnamefont {K.~N.}\ \bibnamefont
  {Cassemiro}}, \bibinfo {author} {\bibfnamefont {V.}~\bibnamefont
  {Poto\ifmmode~\check{c}\else \v{c}\fi{}ek}}, \bibinfo {author} {\bibfnamefont
  {A.}~\bibnamefont {G\'abris}}, \bibinfo {author} {\bibfnamefont
  {I.}~\bibnamefont {Jex}}, \ and\ \bibinfo {author} {\bibfnamefont
  {C.}~\bibnamefont {Silberhorn}},\ }\href {\doibase
  10.1103/PhysRevLett.106.180403} {\bibfield  {journal} {\bibinfo  {journal}
  {Phys. Rev. Lett.}\ }\textbf {\bibinfo {volume} {106}},\ \bibinfo {pages}
  {180403} (\bibinfo {year} {2011})}\BibitemShut {NoStop}%
\bibitem [{\citenamefont {Pandey}\ \emph {et~al.}(2011)\citenamefont {Pandey},
  \citenamefont {Satapathy}, \citenamefont {Meena},\ and\ \citenamefont
  {Ramachandran}}]{Pandey}%
  \BibitemOpen
  \bibfield  {author} {\bibinfo {author} {\bibfnamefont {D.}~\bibnamefont
  {Pandey}}, \bibinfo {author} {\bibfnamefont {N.}~\bibnamefont {Satapathy}},
  \bibinfo {author} {\bibfnamefont {M.~S.}\ \bibnamefont {Meena}}, \ and\
  \bibinfo {author} {\bibfnamefont {H.}~\bibnamefont {Ramachandran}},\ }\href
  {\doibase 10.1103/PhysRevA.84.042322} {\bibfield  {journal} {\bibinfo
  {journal} {Phys. Rev. A}\ }\textbf {\bibinfo {volume} {84}},\ \bibinfo
  {pages} {042322} (\bibinfo {year} {2011})}\BibitemShut {NoStop}%
\bibitem [{\citenamefont {Caruso}\ \emph {et~al.}(2011)\citenamefont {Caruso},
  \citenamefont {Spagnolo}, \citenamefont {Vitelli}, \citenamefont
  {Sciarrino},\ and\ \citenamefont {Plenio}}]{Caruso}%
  \BibitemOpen
  \bibfield  {author} {\bibinfo {author} {\bibfnamefont {F.}~\bibnamefont
  {Caruso}}, \bibinfo {author} {\bibfnamefont {N.}~\bibnamefont {Spagnolo}},
  \bibinfo {author} {\bibfnamefont {C.}~\bibnamefont {Vitelli}}, \bibinfo
  {author} {\bibfnamefont {F.}~\bibnamefont {Sciarrino}}, \ and\ \bibinfo
  {author} {\bibfnamefont {M.~B.}\ \bibnamefont {Plenio}},\ }\href {\doibase
  10.1103/PhysRevA.83.013811} {\bibfield  {journal} {\bibinfo  {journal} {Phys.
  Rev. A}\ }\textbf {\bibinfo {volume} {83}},\ \bibinfo {pages} {013811}
  (\bibinfo {year} {2011})}\BibitemShut {NoStop}%
\bibitem [{\citenamefont {Klinger}\ \emph {et~al.}(2010)\citenamefont
  {Klinger}, \citenamefont {Degenkolb}, \citenamefont {Gemelke}, \citenamefont
  {Soderberg},\ and\ \citenamefont {Chin}}]{Klinger}%
  \BibitemOpen
  \bibfield  {author} {\bibinfo {author} {\bibfnamefont {A.}~\bibnamefont
  {Klinger}}, \bibinfo {author} {\bibfnamefont {S.}~\bibnamefont {Degenkolb}},
  \bibinfo {author} {\bibfnamefont {N.}~\bibnamefont {Gemelke}}, \bibinfo
  {author} {\bibfnamefont {K.-A.~B.}\ \bibnamefont {Soderberg}}, \ and\
  \bibinfo {author} {\bibfnamefont {C.}~\bibnamefont {Chin}},\ }\href {\doibase
  10.1063/1.3274813} {\bibfield  {journal} {\bibinfo  {journal} {Review of
  Scientific Instruments}\ }\textbf {\bibinfo {volume} {81}},\ \bibinfo {eid}
  {013109} (\bibinfo {year} {2010})}\BibitemShut {NoStop}%
\bibitem [{\citenamefont {Li}\ \emph {et~al.}(2005)\citenamefont {Li},
  \citenamefont {Yao}, \citenamefont {Yu}, \citenamefont {Xi},\ and\
  \citenamefont {Chicharo}}]{Li}%
  \BibitemOpen
  \bibfield  {author} {\bibinfo {author} {\bibfnamefont {E.}~\bibnamefont
  {Li}}, \bibinfo {author} {\bibfnamefont {J.}~\bibnamefont {Yao}}, \bibinfo
  {author} {\bibfnamefont {D.}~\bibnamefont {Yu}}, \bibinfo {author}
  {\bibfnamefont {J.}~\bibnamefont {Xi}}, \ and\ \bibinfo {author}
  {\bibfnamefont {J.}~\bibnamefont {Chicharo}},\ }\href {\doibase
  10.1364/OL.30.000189} {\bibfield  {journal} {\bibinfo  {journal} {Opt.
  Lett.}\ }\textbf {\bibinfo {volume} {30}},\ \bibinfo {pages} {189} (\bibinfo
  {year} {2005})}\BibitemShut {NoStop}%
\bibitem [{\citenamefont {Sadgrove}\ and\ \citenamefont
  {Nakagawa}(2011)}]{Sadgrove}%
  \BibitemOpen
  \bibfield  {author} {\bibinfo {author} {\bibfnamefont {M.}~\bibnamefont
  {Sadgrove}}\ and\ \bibinfo {author} {\bibfnamefont {K.}~\bibnamefont
  {Nakagawa}},\ }\href {\doibase 10.1063/1.3655447} {\bibfield  {journal}
  {\bibinfo  {journal} {Review of Scientific Instruments}\ }\textbf {\bibinfo
  {volume} {82}},\ \bibinfo {eid} {113104} (\bibinfo {year}
  {2011})}\BibitemShut {NoStop}%
\bibitem [{\citenamefont {Saleh}\ and\ \citenamefont
  {Teich}(2007)}]{Saleh-Teich}%
  \BibitemOpen
  \bibfield  {author} {\bibinfo {author} {\bibfnamefont {B.~E.~A.}\
  \bibnamefont {Saleh}}\ and\ \bibinfo {author} {\bibfnamefont {M.~C.}\
  \bibnamefont {Teich}},\ }\enquote {\bibinfo {title} {Fundamentals of
  photonics, 2/ed},}\ \ (\bibinfo  {publisher} {Wiley-Interscience},\ \bibinfo
  {year} {2007})\ Chap.\ \bibinfo {chapter} {19. Acousto-Optics}\BibitemShut
  {NoStop}%
\bibitem [{\citenamefont {Baym}(1998)}]{Baym}%
  \BibitemOpen
  \bibfield  {author} {\bibinfo {author} {\bibfnamefont {G.}~\bibnamefont
  {Baym}},\ }\href@noop {} {\bibfield  {journal} {\bibinfo  {journal} {Acta.
  Phys. Polonica}\ }\textbf {\bibinfo {volume} {B29}},\ \bibinfo {pages} {1839}
  (\bibinfo {year} {1998})}\BibitemShut {NoStop}%
%%CITATION = PHRVA,D17,2369;%%
\bibitem [{\citenamefont {Satapathy}\ \emph {et~al.}(2012)\citenamefont
  {Satapathy}, \citenamefont {Pandey}, \citenamefont {Mehta}, \citenamefont
  {Sinha}, \citenamefont {Samuel},\ and\ \citenamefont
  {Ramachandran}}]{Satapathy}%
  \BibitemOpen
  \bibfield  {author} {\bibinfo {author} {\bibfnamefont {N.}~\bibnamefont
  {Satapathy}}, \bibinfo {author} {\bibfnamefont {D.}~\bibnamefont {Pandey}},
  \bibinfo {author} {\bibfnamefont {P.}~\bibnamefont {Mehta}}, \bibinfo
  {author} {\bibfnamefont {S.}~\bibnamefont {Sinha}}, \bibinfo {author}
  {\bibfnamefont {J.}~\bibnamefont {Samuel}}, \ and\ \bibinfo {author}
  {\bibfnamefont {H.}~\bibnamefont {Ramachandran}},\ }\href
  {http://stacks.iop.org/0295-5075/97/i=5/a=50011} {\bibfield  {journal}
  {\bibinfo  {journal} {EPL (Europhysics Letters)}\ }\textbf {\bibinfo {volume}
  {97}},\ \bibinfo {pages} {50011} (\bibinfo {year} {2012})}\BibitemShut
  {NoStop}%
\bibitem [{\citenamefont {Plenio}\ and\ \citenamefont
  {Huelga}(2002)}]{Plenio1}%
  \BibitemOpen
  \bibfield  {author} {\bibinfo {author} {\bibfnamefont {M.~B.}\ \bibnamefont
  {Plenio}}\ and\ \bibinfo {author} {\bibfnamefont {S.~F.}\ \bibnamefont
  {Huelga}},\ }\href {\doibase 10.1103/PhysRevLett.88.197901} {\bibfield
  {journal} {\bibinfo  {journal} {Phys. Rev. Lett.}\ }\textbf {\bibinfo
  {volume} {88}},\ \bibinfo {pages} {197901} (\bibinfo {year}
  {2002})}\BibitemShut {NoStop}%
\bibitem [{\citenamefont {Beige}\ \emph {et~al.}(2000)\citenamefont {Beige},
  \citenamefont {Bose}, \citenamefont {Braun}, \citenamefont {Huelga},
  \citenamefont {Knight}, \citenamefont {Plenio},\ and\ \citenamefont
  {Vedral}}]{Beige}%
  \BibitemOpen
  \bibfield  {author} {\bibinfo {author} {\bibfnamefont {A.}~\bibnamefont
  {Beige}}, \bibinfo {author} {\bibfnamefont {S.}~\bibnamefont {Bose}},
  \bibinfo {author} {\bibfnamefont {D.}~\bibnamefont {Braun}}, \bibinfo
  {author} {\bibfnamefont {S.~F.}\ \bibnamefont {Huelga}}, \bibinfo {author}
  {\bibfnamefont {P.~L.}\ \bibnamefont {Knight}}, \bibinfo {author}
  {\bibfnamefont {M.~B.}\ \bibnamefont {Plenio}}, \ and\ \bibinfo {author}
  {\bibfnamefont {V.}~\bibnamefont {Vedral}},\ }\href {\doibase
  10.1080/09500340008232183} {\bibfield  {journal} {\bibinfo  {journal}
  {Journal of Modern Optics}\ }\textbf {\bibinfo {volume} {47}},\ \bibinfo
  {pages} {2583} (\bibinfo {year} {2000})}\BibitemShut {NoStop}%
\bibitem [{\citenamefont {Caruso}\ \emph {et~al.}(2010)\citenamefont {Caruso},
  \citenamefont {Huelga},\ and\ \citenamefont {Plenio}}]{Caruso2}%
  \BibitemOpen
  \bibfield  {author} {\bibinfo {author} {\bibfnamefont {F.}~\bibnamefont
  {Caruso}}, \bibinfo {author} {\bibfnamefont {S.~F.}\ \bibnamefont {Huelga}},
  \ and\ \bibinfo {author} {\bibfnamefont {M.~B.}\ \bibnamefont {Plenio}},\
  }\href {\doibase 10.1103/PhysRevLett.105.190501} {\bibfield  {journal}
  {\bibinfo  {journal} {Phys. Rev. Lett.}\ }\textbf {\bibinfo {volume} {105}},\
  \bibinfo {pages} {190501} (\bibinfo {year} {2010})}\BibitemShut {NoStop}%
\bibitem [{\citenamefont {Kendon}\ and\ \citenamefont
  {Tregenna}(2003)}]{Kendon}%
  \BibitemOpen
  \bibfield  {author} {\bibinfo {author} {\bibfnamefont {V.}~\bibnamefont
  {Kendon}}\ and\ \bibinfo {author} {\bibfnamefont {B.}~\bibnamefont
  {Tregenna}},\ }\href {\doibase 10.1103/PhysRevA.67.042315} {\bibfield
  {journal} {\bibinfo  {journal} {Phys. Rev. A}\ }\textbf {\bibinfo {volume}
  {67}},\ \bibinfo {pages} {042315} (\bibinfo {year} {2003})}\BibitemShut
  {NoStop}%
\bibitem [{\citenamefont {Mohseni}\ \emph {et~al.}(2008)\citenamefont
  {Mohseni}, \citenamefont {Rebentrost}, \citenamefont {Lloyd},\ and\
  \citenamefont {Aspuru-Guzik}}]{Mohseni}%
  \BibitemOpen
  \bibfield  {author} {\bibinfo {author} {\bibfnamefont {M.}~\bibnamefont
  {Mohseni}}, \bibinfo {author} {\bibfnamefont {P.}~\bibnamefont {Rebentrost}},
  \bibinfo {author} {\bibfnamefont {S.}~\bibnamefont {Lloyd}}, \ and\ \bibinfo
  {author} {\bibfnamefont {A.}~\bibnamefont {Aspuru-Guzik}},\ }\href {\doibase
  10.1063/1.3002335} {\bibfield  {journal} {\bibinfo  {journal} {The Journal of
  Chemical Physics}\ }\textbf {\bibinfo {volume} {129}},\ \bibinfo {eid}
  {174106} (\bibinfo {year} {2008})}\BibitemShut {NoStop}%
\bibitem [{\citenamefont {Plenio}\ and\ \citenamefont {Huelga}(2008)}]{Plenio}%
  \BibitemOpen
  \bibfield  {author} {\bibinfo {author} {\bibfnamefont {M.~B.}\ \bibnamefont
  {Plenio}}\ and\ \bibinfo {author} {\bibfnamefont {S.~F.}\ \bibnamefont
  {Huelga}},\ }\href {http://stacks.iop.org/1367-2630/10/i=11/a=113019}
  {\bibfield  {journal} {\bibinfo  {journal} {New Journal of Physics}\ }\textbf
  {\bibinfo {volume} {10}},\ \bibinfo {pages} {113019} (\bibinfo {year}
  {2008})}\BibitemShut {NoStop}%
\bibitem [{\citenamefont {Caruso}\ \emph {et~al.}(2009)\citenamefont {Caruso},
  \citenamefont {Chin}, \citenamefont {Datta}, \citenamefont {Huelga},\ and\
  \citenamefont {Plenio}}]{Caruso3}%
  \BibitemOpen
  \bibfield  {author} {\bibinfo {author} {\bibfnamefont {F.}~\bibnamefont
  {Caruso}}, \bibinfo {author} {\bibfnamefont {A.~W.}\ \bibnamefont {Chin}},
  \bibinfo {author} {\bibfnamefont {A.}~\bibnamefont {Datta}}, \bibinfo
  {author} {\bibfnamefont {S.~F.}\ \bibnamefont {Huelga}}, \ and\ \bibinfo
  {author} {\bibfnamefont {M.~B.}\ \bibnamefont {Plenio}},\ }\href {\doibase
  10.1063/1.3223548} {\bibfield  {journal} {\bibinfo  {journal} {The Journal of
  Chemical Physics}\ }\textbf {\bibinfo {volume} {131}},\ \bibinfo {eid}
  {105106} (\bibinfo {year} {2009})}\BibitemShut {NoStop}%
\bibitem [{\citenamefont {Ou}\ and\ \citenamefont {Kuang}(2007)}]{Ou}%
  \BibitemOpen
  \bibfield  {author} {\bibinfo {author} {\bibfnamefont {L.-H.}\ \bibnamefont
  {Ou}}\ and\ \bibinfo {author} {\bibfnamefont {L.-M.}\ \bibnamefont {Kuang}},\
  }\href {http://stacks.iop.org/0953-4075/40/i=10/a=017} {\bibfield  {journal}
  {\bibinfo  {journal} {Journal of Physics B: Atomic, Molecular and Optical
  Physics}\ }\textbf {\bibinfo {volume} {40}},\ \bibinfo {pages} {1833}
  (\bibinfo {year} {2007})}\BibitemShut {NoStop}%
\bibitem [{\citenamefont {Bai}\ and\ \citenamefont {Han}(2007)}]{Bai}%
  \BibitemOpen
  \bibfield  {author} {\bibinfo {author} {\bibfnamefont {Y.}~\bibnamefont
  {Bai}}\ and\ \bibinfo {author} {\bibfnamefont {S.}~\bibnamefont {Han}},\
  }\href {\doibase 10.1103/PhysRevA.76.043828} {\bibfield  {journal} {\bibinfo
  {journal} {Phys. Rev. A}\ }\textbf {\bibinfo {volume} {76}},\ \bibinfo
  {pages} {043828} (\bibinfo {year} {2007})}\BibitemShut {NoStop}%
\bibitem [{\citenamefont {Agafonov}\ \emph {et~al.}(2008)\citenamefont
  {Agafonov}, \citenamefont {Chekhova}, \citenamefont {Iskhakov},\ and\
  \citenamefont {Penin}}]{Agafanov}%
  \BibitemOpen
  \bibfield  {author} {\bibinfo {author} {\bibfnamefont {I.~N.}\ \bibnamefont
  {Agafonov}}, \bibinfo {author} {\bibfnamefont {M.~V.}\ \bibnamefont
  {Chekhova}}, \bibinfo {author} {\bibfnamefont {T.~S.}\ \bibnamefont
  {Iskhakov}}, \ and\ \bibinfo {author} {\bibfnamefont {A.~N.}\ \bibnamefont
  {Penin}},\ }\href {\doibase 10.1103/PhysRevA.77.053801} {\bibfield  {journal}
  {\bibinfo  {journal} {Phys. Rev. A}\ }\textbf {\bibinfo {volume} {77}},\
  \bibinfo {pages} {053801} (\bibinfo {year} {2008})}\BibitemShut {NoStop}%
\bibitem [{\citenamefont {Cao}\ \emph {et~al.}(2008)\citenamefont {Cao},
  \citenamefont {Xiong}, \citenamefont {Zhang}, \citenamefont {Lin},
  \citenamefont {Gao},\ and\ \citenamefont {Wang}}]{Cao}%
  \BibitemOpen
  \bibfield  {author} {\bibinfo {author} {\bibfnamefont {D.-Z.}\ \bibnamefont
  {Cao}}, \bibinfo {author} {\bibfnamefont {J.}~\bibnamefont {Xiong}}, \bibinfo
  {author} {\bibfnamefont {S.-H.}\ \bibnamefont {Zhang}}, \bibinfo {author}
  {\bibfnamefont {L.-F.}\ \bibnamefont {Lin}}, \bibinfo {author} {\bibfnamefont
  {L.}~\bibnamefont {Gao}}, \ and\ \bibinfo {author} {\bibfnamefont
  {K.}~\bibnamefont {Wang}},\ }\href {\doibase 10.1063/1.2919719} {\bibfield
  {journal} {\bibinfo  {journal} {Applied Physics Letters}\ }\textbf {\bibinfo
  {volume} {92}},\ \bibinfo {eid} {201102} (\bibinfo {year}
  {2008})}\BibitemShut {NoStop}%
\bibitem [{\citenamefont {Pandey}\ \emph {et~al.}()\citenamefont {Pandey},
  \citenamefont {Satapathy}, \citenamefont {Suryabrahmam}, \citenamefont
  {Ivan},\ and\ \citenamefont {Ramachandran}}]{Pandey2}%
  \BibitemOpen
  \bibfield  {author} {\bibinfo {author} {\bibfnamefont {D.}~\bibnamefont
  {Pandey}}, \bibinfo {author} {\bibfnamefont {N.}~\bibnamefont {Satapathy}},
  \bibinfo {author} {\bibfnamefont {B.}~\bibnamefont {Suryabrahmam}}, \bibinfo
  {author} {\bibfnamefont {J.~S.}\ \bibnamefont {Ivan}}, \ and\ \bibinfo
  {author} {\bibfnamefont {H.}~\bibnamefont {Ramachandran}},\ }\href@noop {}
  {\bibinfo  {journal} {to be published}\ }\BibitemShut {NoStop}%
\end{thebibliography}
\end{document}